\documentclass[twocolumn,aps,prd]{revtex4}
 \usepackage{graphicx,amssymb}
 \usepackage[utf8]{inputenc} 
 
\usepackage[urlcolor=blue,colorlinks,breaklinks=true]{hyperref}
\usepackage[anythingbreaks]{breakurl}

\PassOptionsToPackage{obeyspaces,spaces,hyphens}{url}
 
 \usepackage{amsmath}

\begin{document}
 	\def\half{{1\over2}}
 	\def\shalf{\textstyle{{1\over2}}}
 	
 	\newcommand\lsim{\mathrel{\rlap{\lower4pt\hbox{\hskip1pt$\sim$}}
 			\raise1pt\hbox{$<$}}}
 	\newcommand\gsim{\mathrel{\rlap{\lower4pt\hbox{\hskip1pt$\sim$}}
 			\raise1pt\hbox{$>$}}}

 	\title{Particle creation and decay in nonminimally coupled models of gravity}
 	
 	\author{R.P.L. Azevedo}
 	\email[Electronic address: ]{rplazevedo@fc.up.pt}
 	\affiliation{Instituto de Astrof\'{\i}sica e Ci\^encias do Espa{\c c}o, Universidade do Porto, CAUP, Rua das Estrelas, PT4150-762 Porto, Portugal}
 	\affiliation{Centro de Astrof\'{\i}sica da Universidade do Porto, Rua das Estrelas, PT4150-762 Porto, Portugal}
 	\affiliation{Departamento de F\'{\i}sica e Astronomia, Faculdade de Ci\^encias, Universidade do Porto, Rua do Campo Alegre 687, PT4169-007 Porto, Portugal}
 	
 	\author{P.P. Avelino}
 	\email[Electronic address: ]{pedro.avelino@astro.up.pt}
 	\affiliation{Instituto de Astrof\'{\i}sica e Ci\^encias do Espa{\c c}o, Universidade do Porto, CAUP, Rua das Estrelas, PT4150-762 Porto, Portugal}
 	\affiliation{Centro de Astrof\'{\i}sica da Universidade do Porto, Rua das Estrelas, PT4150-762 Porto, Portugal}
 	\affiliation{Departamento de F\'{\i}sica e Astronomia, Faculdade de Ci\^encias, Universidade do Porto, Rua do Campo Alegre 687, PT4169-007 Porto, Portugal}
 	
 	\date{\today}
 	\begin{abstract}
 		
 		In extended models of gravity a nonminimal coupling to matter has been assumed to lead to irreversible particle creation. In this paper we challenge this assumption. We argue that a non-minimal coupling of the matter and gravitational sectors results in a change in particle-momentum on a cosmological timescale, irrespective of particle creation or decay. We further argue that particle creation or decay associated with a non-minimal coupling to gravity could only happen as a result of significant deviations from a homogeneous Friedmann-Lemaître-Robertson-Walker geometry on microscopic scales, and provide a phenomenological description of the impact of particle creation or decay on the cosmological evolution of the density of the matter fields. 
 		
 	\end{abstract}
 	\maketitle
 	
 	\section{Introduction}
 	\label{sec:intr}
 	
 	Despite its success, modern cosmology is faced with great challenges, including the determination of the origin of the accelerated expansion of the Universe, and of the detailed dynamics of galaxies and clusters \cite{Bertone2005,Carroll2001}. Within the framework of General Relativity (GR), one usually requires dark matter and dark energy to dominate the energy density of the Universe to explain the large scale dynamics of the Universe. Alternatively, one can forgo this exotic dark energy component (or even dark matter --- see, however, \cite{Clowe2004,Markevitch2004}) and instead look for extensions of GR that more naturally feature phases of accelerated expansion, such as theories with additional fields, theories with more complex geometric terms and theories featuring a nonminimal coupling (NMC) between geometry and matter, such as $f(R,\mathcal{L}_m)$ theories, where $R$ and $\mathcal{L}_m$ are respectively the Ricci scalar and the Lagrangian density of the matter fields \cite{Nojiri2004,Allemandi:2005qs,Bertolami2007,Sotiriou2008,Harko2010,Harko2011}.
 	
 	An important feature of NMC theories is that energy-momentum is not usually covariantly conserved, as a consequence of the matter Lagrangian featuring explicitly in the equations of motion. This leads to significant consequences, in particular in a cosmological context \cite{Avelino2018, Azevedo2018a}. The predictions of NMC theories  are crucially dependent on the Lagrangian of the matter fields, and it is therefore imperative that the matter fields  are appropriately described. In previous work $\mathcal{L}_m=-\rho$ or $\mathcal{L}_m=p$ have been suggested as the on-shell Lagrangian of a perfect fluid with proper energy density $\rho$ and pressure $p$ \cite{Nesseris2009,Bertolami2012,Azizi:2014qsa,Ribeiro2014,Silva2018}. However, it has recently been shown that the correct on-shell  Lagrangian for a fluid composed of solitonic particles of fixed rest mass and structure is given by the trace of the energy-momentum tensor of the fluid $\mathcal{L}_m = T = 3p-\rho$ \cite{Avelino2018a,Avelino2018}. While this description does not apply to dark energy (or to any fluid with an equation of state parameter outside the interval $0\leq w \leq 1/3$), it is expected to be a good approximation in the case of baryonic matter, dark matter and photons (the zero rest mass limit has been considered in the case of photons). So far it has been used in the derivation of stringent constraints on NMC gravity originating from Cosmic Microwave Background (CMB) and Big-Bang Nucleosynthesis observations \cite{Avelino2018, Azevedo2018a}.
 	
 	Energy-momentum non-conservation in NMC theories has been suggested to be associated with gravitationally induced particle creation, following a thermodynamic analysis, which assumed the Lagrangian $\mathcal{L}_m=-\rho$ for a perfect fluid \cite{Harko2015}. In this paper, we extend this analysis arguing that rather than particle creation, the use of the correct Lagrangian implies a change of particle momentum on cosmological timescales. We also provide a phenomenological description of particle creation/decay associated to the presence of significant perturbations to the spacetime geometry on microscopic scales.
 	
 	Throughout this paper we use fundamental units such that $c=\hbar=k_B=1$. Here $c$ is the value of the speed of light in vacuum, $\hbar$ is the reduced Planck constant, and $k_B$ is the Boltzmann constant. We adopt the metric signature $(-,+,+,+)$, and the Einstein summation convention will be used as usual.
 	
 	\section{Non-minimally coupled gravity}
 	\label{sec:model}
 	
 	Many forms of nonminimal coupling models have been proposed in the literature, some including more complex geometrical terms like $f(R,T,R_{\mu\nu}T^{\mu\nu})$ theories \cite{Odintsov2013,Haghani2013}. Here, we shall consider a model inspired by $f(R)$ theories due to its fairly simple form, broad explanatory power, and for being able to avoid the Ostrogradsky and Dolgov-Kawasaki instabilities \cite{Bertolami2009,Ayuso2015}. It is described by the action
 	\begin{equation}
 	\label{eq:action}
 	S=\int \sqrt{-g} \left[\kappa f_1(R) + f_2(R)\mathcal{L}_m\right]\,,
 	\end{equation}
 	where $\kappa=(16\pi G)^{-1}$, $G$ is Newton's gravitational constant, $g$ is the determinant of the metric $g_{\mu\nu}$, $\mathcal{L}_m$ is the Lagrangian of the matter fields, and $f_1(R)$ and $f_2(R)$ are generic functions of the Ricci scalar $R$. GR is recovered if $f_1(R)=R$ and $f_2(R)=1$. Extremizing the action with respect to the metric one obtains the equations of motion of the gravitational field
 	\begin{equation}
 	\label{eq:eqmotion}
 	F G_{\mu\nu}=\half f_2 T_{\mu\nu}+\Delta_{\mu\nu}F+\half\kappa f_1 g_{\mu\nu}-\half RFg_{\mu\nu}\, ,
 	\end{equation}
 	where $G_{\mu\nu}=R_{\mu\nu}-\shalf g_{\mu\nu} R $ is the Einstein tensor, $R_{\mu\nu}$ is the Ricci tensor, $\Delta_{\mu \nu} \equiv \nabla_\mu \nabla_\nu - g_{\mu \nu} \Box$, $\Box \equiv \nabla^\mu \nabla_\mu$,  
 	\begin{equation}
 	\label{eq:F}
 	F=\kappa f'_1(R)+f'_2(R)\mathcal{L}_m\,,
 	\end{equation}
 	a prime denotes a derivative with respect to the Ricci scalar, and the energy-momentum tensor has the usual form
 	\begin{equation}
 	\label{eq:energymom}
 	T_{\mu\nu}=-{2\over \sqrt{-g}}{\delta(\sqrt{-g}\mathcal{L}_m)\over \delta g^{\mu\nu}}\,.
 	\end{equation}
 	
 	Taking the covariant derivative of Eq.~\eqref{eq:eqmotion} and using the Bianchi identities one obtains the following relation in lieu of the usual energy-momentum conservation equation
 	\begin{equation}
 	\label{eq:noncons}
 	\nabla^\mu T_{\mu\nu}={f'_2\over f_2}(g_{\mu\nu}\mathcal{L}_m-T_{\mu\nu})\nabla^\mu R\, .
 	\end{equation}
 	Equation~\eqref{eq:noncons} implies that the form of the matter Lagrangian directly affects not only energy-momentum conservation, but also particle motion \cite{Bertolami2008a,Avelino2018}. In fact, introducing  the projection operator $h_{\mu\nu}=g_{\mu\nu}+u_{\mu}u_{\nu}$, where $u^\mu$ is the four-velocity of the fluid, results in the non-geodesic equation for the motion of a perfect fluid element
 	\begin{equation}
 	\label{eq:geodesic}
 	{du^{\mu}\over ds}+\Gamma^\mu_{\alpha\beta} u^\alpha u^\beta = f^\mu \, ,
 	\end{equation}
 	where $f^\mu$ is an extra force given by
 	\begin{equation}
 	\label{eq:extraforce}
 	f^\mu={1\over \rho+p}\left[{f'_2\over f_2}(\mathcal{L}_m-p)\nabla_\nu R + \nabla_\nu p\right]h^{\mu\nu} \, ,
 	\end{equation}
 	and $\rho$ and $p$ are respectively the proper energy density and pressure of a perfect fluid with energy-momentum tensor
 	\begin{equation}
 	\label{eq:pfemt}
 	T^{\mu\nu}=(\rho+p)u^\mu u^\nu + p g^{\mu\nu}\,.
 	\end{equation}
 	
 	In previous work \cite{Avelino2018a,Avelino2018} it was determined that the on-shell Lagrangian of a perfect fluid composed of non-interacting particles with fixed mass and structure, i.e. solitons, is given by
 	\begin{equation}
 	\label{eq:lagrangian}
 	\mathcal{L}_m=T=3p-\rho\,,
 	\end{equation}
 	where $T=T^\mu_{~~\mu}$ is the trace of the energy-momentum tensor. The particular structure of the particles is not relevant for this derivation. In the derivation of the Lagrangian it was assumed that particles described by this Lagrangian can neither decay (or conversely, be created) nor experience fundamental changes to their structure or mass as a result of the NMC to gravity \cite{Uzan2011,Copeland2007}.

 	A flat homogeneous and isotropic universe is described by the flat Friedmann-Lemaître-Robertson-Walker (FLRW) metric with line element
 	\begin{equation}
 	\label{eq:metric}
 	ds^2=-dt^2+a^2(t)\left[dx^2 + dy^2 +dz^2\right]\,,
 	\end{equation} 
 	where $a(t)$ is the scale factor, $t$ is the cosmic time, and $x$, $y$, and $z$ are Cartesian comoving coordinates.
 	
 	The dynamics of $p$-branes in flat $N+1$-dimensional FLRW universes has been studied in detail in \cite{Sousa2011a,Sousa2011b} (see also \cite{Avelino2016}). There it has been shown that the evolution of the velocity $v$ of a soliton in a flat $3+1$-dimensional FLRW spacetime (ignoring interactions other than gravitational) is given by
 	\begin{equation}
 	\label{eq:dotvel}
 	{\dot v} +3\left( H + \frac{{\dot f}_2}{f_2} \right) (1-v^2) v =0 \,.
 	\end{equation}
 	where $H\equiv \dot{a}/a$ is the Hubble parameter, and a dot represents a derivative with respect to the cosmic time. Hence, the momentum of such a particle evolves as
 	\begin{equation}
 	m \gamma v \propto (a f_2)^{-1}\,, \label{eq:momev}
 	\end{equation}
 	where $\gamma \equiv (1-v^2)^{-1/2}$.
 	
 	The 0th component of Eq.~\eqref{eq:noncons} is given by
 	\begin{equation}
 	\label{eq:densityevolution}
 	{\dot \rho} +  3 H (\rho +p) =  -({\mathcal L}_m  + \rho) \frac{\dot f_2}{f_2} \, .
 	\end{equation} 
 	Taking into account that the proper pressure of the fluid is given by $p=\rho   v^2/3$ (assuming, for simplicity, that $v$ is the same for all particles) and requiring that the number of particles per comoving volume be conserved, or equivalently that $\rho \propto \gamma a^{-3}$, it is straightforward to show that the consistency between Eqs. \eqref{eq:momev} and \eqref{eq:densityevolution} implies that the matter Lagrangian is indeed given by Eq.~\eqref{eq:lagrangian} (see \cite{Avelino2018a,Avelino2018} for alternative derivations of the same result).
 	
 	\section{Energy non-conservation and particle creation/decay}\label{sec:thermo}
 	
 	Particle creation or decay via a NMC to gravity would require significant perturbations to the FLRW geometry on the relevant microscopic scales, since the FLRW metric is essentially Minkowskian on such scales. The constraints on gravity on microscopic scales are extremely weak and it is possible to construct viable modified theories of gravity in which the gravitational interaction on such scales is significantly enhanced with respect to general relativity (see, for example, \cite{Avelino2012a,Avelino2012}). However, these small scale perturbations have not been considered in the derivation of Eq.~\eqref{eq:densityevolution} and have not been explicitly taken into account in previous works when considering particle creation or decay via a NMC to gravity. Consequently, the only consistent interpretation for the change to the evolution of the energy density of a fluid made of soliton-like particles associated to the term on the right-hand-side of Eq.~\eqref{eq:densityevolution} is the modification to the evolution of the linear momentum of such particles described by Eq.~\eqref{eq:momev}. Here, we shall start by considering the thermodynamics of a homogeneous and isotropic universe in the absence of significant small-scale perturbations, and then describe phenomenologically the case in which microscopic perturbations to the FLRW geometry result in particle creation or decay.
 	
 	\subsection{Perfect fluid with $\mathcal{L}_m=T$}
 	
 	In order to study the implications of the usual energy-momentum tensor conservation law, we shall consider the thermodynamics of a universe filled with a perfect fluid, in the presence of a NMC between geometry and matter described by the action given in Eq.~\eqref{eq:action}. We start by treating the Universe as a system where the number of particles per comoving volume is conserved, for which the first law of thermodynamics takes the form
 	\begin{equation}
 	\label{eq:conservenergy}
 	d (\rho a^3)= dQ_{\rm NMC}-pd(a^3) \, ,
 	\end{equation} 
 	where $dQ_{\rm NMC}$ is the ``heat'' received by the system over the interval of time $dt$ due to the NMC between the gravitational and the matter fields. As previously mentioned, in the literature \cite{Prigogine1988, Prigogine1989, Lima2014, Harko2015}, an adiabatic expansion ($dQ/dt=0$) is usually considered and an extra term, associated to particle creation due to the NMC between the gravitation and matter fields, is added to Eq.~\eqref{eq:conservenergy}. However, and given the lack of alterations to the microscopic geometry responsible for these terms, we are left with associating the NMC with the non-adiabaticity of the expansion.
 	
 	Eq.~\eqref{eq:conservenergy} may be rewritten as
 	\begin{equation}
 	\label{eq:heat1}
 	\dot{\rho}+3H(\rho+p)={{\dot{Q}}_{\rm NMC}\over a^3} \, .
 	\end{equation}
 	Using Eq.~\eqref{eq:densityevolution}, one obtains the ``heat'' transfer rate with $\mathcal{L}_m=3p-\rho$
 	\begin{eqnarray}
 	\label{eq:heat_transf}
 	{\dot{Q}}_{\rm NMC}&=&-({\mathcal L}_m  + \rho) a^3 \frac{\dot f_2}{f_2}=-3p a^3\frac{\dot f_2}{f_2} \nonumber \\
 	&=&-\rho v^2 a^3\frac{\dot f_2}{f_2} \, .
 	\end{eqnarray}
 	This implies that for non-relativistic matter ($v\ll1$), such as baryons and cold dark matter, ${\dot{Q}}_{\rm NMC}\sim 0$ so that the usual energy-momentum conservation approximately holds. On the other hand, relativistic matter and photons are strongly impacted by this energy-momentum transfer which is responsible for a new source of spectral distortion ($n$-type spectral distortions) of the CMB power spectrum already discussed in Ref.~\cite{Avelino2018}.
 	
 	\subsection{Particle creation/decay and effective Lagrangians}
 	
 	Here, we consider the possibility that the perturbations to the FLRW geometry on microscopic scales may be responsible for particle creation or decay. Discussing particle creation/decay with the matter Lagrangian $\mathcal{L}_m=T$ in great detail would of course require a microscopic description of the particle structure, which we leave purposefully generic, and its interaction with gravity on microscopic scales. While such analysis is beyond the scope of the present paper, we can treat particle creation/decay phenomenologically, by introducing a modification to the energy-momentum conservation equation. If particle number is not conserved due to the NMC, an additional term, associated to particle creation/decay, should therefore be added to the right-hand side of Eq.~\eqref{eq:densityevolution} 
 	\begin{equation}
 	\label{eq:densitycreation}
 	 {\dot \rho} +  3 H (\rho +p) =  -({\mathcal L}_m  + \rho) \frac{\dot f_2}{f_2} -\mathcal{L}_\Gamma \frac{\dot f_2}{f_2} \, .
 	\end{equation}
 	Note that $\mathcal{L}_\Gamma$ is not a true Lagrangian, but rather a phenomenological term associated to the effect of the NMC between matter and gravity on microscopic scales. If the mass and structure of the particles does not change due to the NMC to gravity, except (almost) instantaneous particle creation or decay, the Lagrangian of the perfect fluid is still described by $\mathcal{L}_m=T$ (we also allow for almost instantaneous scattering events which do not have an impact in the form of the perfect fluid Lagrangian). Hence, Eq.~\eqref{eq:momev} still describes the cosmological contribution to the evolution of the linear-momentum of the particles. Equation~\eqref{eq:densitycreation} may then be rewritten as
 	\begin{equation}
	\label{eq:densitycreation2}
	{\dot \rho} +  3 H (\rho +p) =  -({\mathcal L}_{eff}  + \rho) \frac{\dot f_2}{f_2} \, ,
	\end{equation} 	
 	where
 	\begin{equation}
 	\label{eq:efflagrangian}
 	\mathcal{L}_{eff} = \mathcal{L}_m + \mathcal{L}_\Gamma \, .
 	\end{equation}
 	
 	In this case Eq.~\eqref{eq:conservenergy} is changed to \cite{Prigogine1989}
 	\begin{equation}
 	\label{eq:conservenergycreation}
 	d (\rho a^3)= dQ_{\rm NMC}-pd(a^3) + {h\over n}d (n a^3) \, ,
 	\end{equation} 
 	where $n$ is the particle number density and $h=\rho+p$ is the enthalpy per unit volume. For simplicity, we have also implicitly assumed that all particles are identical and that the corresponding perfect fluid is always in thermodynamic equilibrium. This is a natural assumption if the rate of particle creation/decay is much smaller than the particle scattering rate, a case in which thermalization, following particle creation/decay occurs (almost) instantaneously.

Equation~\eqref{eq:conservenergycreation} may be rewritten as as
 	\begin{equation}
 	\label{eq:creation}
 	\dot{\rho}+3H(\rho+p)={{\dot{Q}}_{\rm NMC}\over a^3} + {h\over n}(\dot{n}+3Hn) \, ,
 	\end{equation}
 	and using Eq.~\eqref{eq:densitycreation2} one finds that
 	\begin{equation}
 	\label{eq:creation2}
 	{{\dot{Q}}_{\rm NMC}\over a^3} + {h\over n}(\dot{n}+3Hn)=-(\mathcal{L}_{eff}+\rho) \frac{\dot f_2}{f_2} \, .
 	\end{equation}
 	Equations~\eqref{eq:heat_transf}, \eqref{eq:efflagrangian} and \eqref{eq:creation2} also imply that
 	\begin{equation}
 	\label{eq:creation3}
 	{\rho+p\over n}(\dot{n}+3Hn)=-\mathcal{L}_{\Gamma} \frac{\dot f_2}{f_2}\, .
 	\end{equation}
 	Introducing the particle creation/decay rate 
 	\begin{equation}
 	\label{eq:gamma}
 	\Gamma  = {\dot{n}\over n}+3H \, ,
 	\end{equation}
 	and using Eq.~\eqref{eq:creation3} one obtains
 	\begin{equation}
 	\label{eq:creation4}
 	\Gamma= -{\mathcal{L}_\Gamma\over \rho+p}\frac{\dot f_2}{f_2}\, .
 	\end{equation}
 		
 	 Alternatively, particle creation/decay may be described as an extra effective creation/decay pressure $p_\Gamma$ of the perfect fluid that must be included in the continuity equation \cite{Prigogine1988}
 	\begin{equation}
 	\label{eq:conteqpres}
 	\dot{\rho} +3H(\rho+p+p_\Gamma)= -({\mathcal L}_{m}  + \rho) \frac{\dot f_2}{f_2} \, ,
 	\end{equation}
	where
	\begin{equation}
	\label{eq:creationpressure}
 	p_\Gamma = {\mathcal{L}_{\Gamma}\over 3H} {\dot{f}_2\over f_2}\,
 	\end{equation}
 	may be obtained from Eq. \eqref{eq:densitycreation2}.
 	
	We have argued that the correct form of the Lagrangian of a perfect fluid composed of solitonic particles is $\mathcal{L}_m=T$, even in the presence of (almost) instantaneous particle scattering and/or particle creation/decay, and when $\mathcal{L}_{eff}=\mathcal{L}_m$, one trivially recovers the results of the previous subsection. Nevertheless, one may ask whether or not the Lagrangians suggested in previous work to describe such a perfect fluid could play the role of effective Lagrangians. Let us then consider the particular cases with $\mathcal{L}_{eff}=-\rho$ and $\mathcal{L}_{eff}=p$. 
	
	If $\mathcal{L}_{eff}=-\rho$ then 
	\begin{equation}
	\label{eq:Lgammarho}
	\mathcal{L}_\Gamma = \mathcal{L}_{eff} - \mathcal{L}_m = -3p \, ,
	\end{equation}
	where we have used Eq. \eqref{eq:efflagrangian} and taken into account that $\mathcal{L}_m=T=3p-\rho$. Hence, in this case 	
	\begin{equation}
 	\label{eq:creationpressurerho}
	p_\Gamma = -{p\over H} {\dot{f}_2\over f_2}\, ,
	\end{equation}
	and there is a particle creation/decay rate given by
	\begin{equation}
	\label{eq:gammarho}
	\Gamma= {3p\over \rho+p}\frac{\dot f_2}{f_2}\, .
	\end{equation}
	Notably, if $\mathcal{L}_{eff}=-\rho$ the standard conservation equation for the energy density is recovered.
	
	If $\mathcal{L}_{eff}=p$ then
 	\begin{equation}
 	\label{eq:Lgammap}
 	\mathcal{L}_\Gamma = \rho-2p \, .
 	\end{equation}
 	In this case, the effective pressure is equal to
 	\begin{equation}
 	\label{eq:creationpressurep}
 	p_\Gamma = {\rho-2p\over 3H} {\dot{f}_2\over f_2} \, ,
 	\end{equation}
 	and the particle creation/decay rate is
 	\begin{equation}
 	\label{eq:gammap}
 	\Gamma= -{\rho-2p\over \rho+p}\frac{\dot f_2}{f_2}\, .
 	\end{equation}
 	Note that if ${\mathcal L}_{eff}=p$ the standard evolution equation for the density is not recovered, unless $p=-\rho$.

	In both cases, $\mathcal{L}_{eff}=-\rho$ and $\mathcal{L}_{eff}=p$, the particle creation/decay rate $\Gamma$ would not in general be a constant. Rather than depending on the particle properties and on the way these are affected by the NMC to gravity on microscopic scales, for a given choice of the function $f_2$ the evolution of $\Gamma$ given in Eqs. \eqref{eq:gammarho} and \eqref{eq:gammap} would depend essentially on the cosmology and on the macroscopic properties of the fluid. As discussed before, the FLRW metric is essentially Minkowskian on microscopic scales relevant to particle creation/decay. Consequently, one should not expect such a cosmological dependence of the particle creation/decay rate $\Gamma$, which questions the relevance of the effective Lagrangians $\mathcal{L}_{eff}=-\rho$ and $\mathcal{L}_{eff}=p$. 	
 	
 	\section{Conclusions}\label{sec:conc}

In this work we challenged the assumption that the NMC between geometry and the matter fields might be responsible for particle creation/decay in the absence of significant perturbations to the FLRW metric on microscopic scales. We have argued that there is only one consistent interpretation for the modification to the evolution of the energy density of a fluid made of soliton-like particles associated to the the NMC between the gravitational and the matter fields in a FLRW universe: a change in particle-momentum on a cosmological timescale (rather than particle creation or decay). We have considered the possibility that perturbations to the FLRW geometry on microscopic scales, eventually in association to significant extensions to the NMC theory of gravity studied in the present paper, may be responsible for particle creation or decay. We have also have provided a phenomenological description of particle creation/decay by defining an ``effective Lagrangian'' which incorporates these effects.
 	
 	\begin{acknowledgments}
 		R.P.L.A. was supported by the Funda{\c c}\~ao para a Ci\^encia e Tecnologia (FCT, Portugal) grant SFRH/BD/132546/2017. P.P.A. acknowledges the support by FEDER—Fundo Europeu de Desenvolvimento Regional funds through the COMPETE 2020—Operational Programme for Competitiveness and Internationalisation (POCI), and by Portuguese funds through FCT - Fundação para a Ciência e a Tecnologia in the framework of the project POCI-01-0145-FEDER-031938. Funding of this work has also been provided by the FCT grant UID/FIS/04434/2013. This paper benefited from the participation of the authors on the COST action CA15117 (CANTATA), supported by COST (European Cooperation in Science and Technology).
 	\end{acknowledgments}
 	
 	\bibliography{NMC_Part_Creation}
 	
 \end{document}